\documentclass[aps,pra,reprint,amsmath,amssymb,graphicx,longbibliography]{revtex4-2}

\usepackage{bm}

\usepackage[normalem]{ulem}
\usepackage[colorlinks=true,allcolors=blue,bookmarks=false,pdfusetitle]{hyperref}
\usepackage{url}
\usepackage{xcolor}
\usepackage{graphicx}
\usepackage{breqn}
\usepackage{braket}
\makeatletter
\let\cat@comma@active\@empty
\makeatother

\begin{document}
%\nocite{*}

\title{The Theory of Generalised Hydrodynamics for the One-Dimensional Bose Gas}

\author{Matthew L. Kerr}
\affiliation{School of Mathematics and Physics, University of Queensland, Brisbane,  Queensland 4072, Australia}
\author{Karen V. Kheruntsyan*}
\affiliation{School of Mathematics and Physics, University of Queensland, Brisbane,  Queensland 4072, Australia}

\date{\today{}}

\begin{abstract}
\noindent This article reviews the recent developments in the theory of generalised hydrodynamics (GHD) with emphasis on the repulsive one-dimensional Bose gas. We discuss the implications of GHD on the mechanisms of thermalisation in integrable quantum many-body systems as well as its ability to describe far-from-equilibrium behaviour of integrable and near-integrable systems in a variety of quantum quench scenarios. We outline the experimental tests of GHD in cold-atom gases and its benchmarks with other microscopic theoretical approaches. Finally, we offer some perspectives on the future direction of the development of GHD.

~


~

\noindent{*} Corresponding author: karen.kheruntsyan@uq.edu.au
\end{abstract}

\maketitle

\section{Introduction}

The study of non-equilibrium phenomena in isolated quantum systems has received much attention in recent years \cite{Cazalilla_2010,Polkovnikov2011, eisert2015quantum, PhysRevLett.127.130601}. 
A fruitful avenue into non-equilibrium dynamics has emerged from the study of integrable systems. These systems are characterised by a large number of conserved quantities and include several paradigmatic models, such as the Lieb-Liniger model describing a gas of one-dimensional bosons \cite{Lieb-Liniger-I, Lieb-Liniger-II}, and the Hubbard model describing electrons in a solid \cite{hubbard}. Integrability is known to fundamentally affect the dynamics of isolated quantum systems in contrast to that of generic non-integrable systems \cite{eisert2015quantum, Gogolin_2016, Polkovnikov2011, PhysRevLett.103.100403}. The most prominent consequence of this is the preclusion of thermalisation, whereby a system does not relax to a thermal state. Rather, integrable systems will typically relax to non-thermal, equilibrium states, described by the so-called generalised Gibbs ensemble \cite{Rigol2007,rigol2008thermalization, experimental-GGE}. This has been demonstrated in various models and theories, such as conformal field theories \cite{PhysRevLett.96.136801, calabrese2016quantum}, the repulsive Lieb-Liniger gas \cite{caux2016quench, PhysRevA.89.033601}, the quantum Ising chain in a transverse field \cite{Calabrese2011Quantum,Calabrese2012QuantumI,Calabrese2012QuantumII,Cazalilla2012Thermalization,essler2016quench}, and in the spin-1/2 anisotropic Heisenberg ($XXZ$) spin chain \cite{Pozsgay2014Correlations,ilievski2016quasilocal}.

In classical systems, thermalisation can be well understood in terms of dynamical chaos and the notions of ergodicity and mixing \cite{loskutov2007dynamical} in which a system explores its phase space uniformly and densely for almost all initial conditions. This enables a description of the system via classical statistical mechanics. However, the mechanisms of thermalisation for isolated quantum systems are not well-established. Understanding these mechanisms has been the subject of many theoretical and experimental investigations over the past two decades.

An indispensable protocol to induce non-equilibrium scenarios in isolated quantum systems is the so-called quantum quench \cite{PhysRevLett.96.136801, rigol2008thermalization, mitra2018quantum} where an initial equilibrium state of a many-body Hamiltonian is suddenly made to evolve unitarily under another Hamiltonian. These quench scenarios constitute a vital outpost to address the lack of thermalisation in integrable quantum systems. However, as is the case for any quantum system, a direct theoretical approach with exact microscopic calculations of physically relevant observables involves the diagonalisation of the many-body Hamiltonian associated with the system. The dimension of such a many-body Hilbert space notoriously scales exponentially with the total number of particles. This often renders direct computational approaches intractable beyond more than a dozen particles. One can transcend this exponentially large space of quantum states by instead employing an aggregated, coarse-grained description of the system with a reduced number of degrees of freedom. One of the most prominent and successful such frameworks is hydrodynamics. 

At the heart of any hydrodynamic theory is a set of conservation laws that govern a specific model's dynamical degrees of freedom over vast space and time scales. These quantities typically include energy, momentum, and particle number density. Hydrodynamics has been particularly successful as an effective theory for the emergent dynamical behaviour of a variety of interacting many-body systems. For instance, Landau's celebrated two-fluid model of superfluid helium \cite{PhysRev.60.356}, the description of electron currents in graphene \cite{levitov2016electron}, and in the magneto-hydrodynamic description of electrically conducting fluids \cite{moreau1990magnetohydrodynamics}.

Hydrodynamic theories are a valuable resource to address the mechanisms of thermalisation in isolated integrable quantum systems. Here our interest lies in the rich many-body physics of one-dimensional integrable systems, which, besides being a mathematical curiosity, are also relevant to ultracold Bose and Fermi gases in highly anisotropic traps, and to other low-dimensional condensed matter systems such as superconducting nanowires \cite{applications} and 1D polariton condensates \cite{PhysRevB.83.144513}. In particular, this review focuses on the one-dimensional Bose gas. This is motivated by the high degree of experimental control over system parameters and dynamics, owing to unprecedented advancements over the past twenty years \cite{RevModPhys.80.885,bouchoule2011atom}. 

By utilising the exact integrability of the one-dimensional Bose gas in the uniform limit, its equilibrium thermodynamic properties can be derived exactly using Yang-Yang thermodynamics \cite{yang1969thermodynamics}. One can then use these thermodynamic properties (more specifically, the thermodynamic equation of state for the pressure of the gas) in the equations of classical hydrodynamics that govern large-scale dynamics of the system.
This approach has been dubbed ``conventional hydrodynamics'' (CHD) and provides an excellent description of the collective excitations of 1D bosons driven out of equilibrium in certain cases \cite{Menotti-Stringari-2002,Hu2014,DeRosi2015,DeRosi2016,bouchoule-hydro-2016}. The equations of CHD take the form of Euler hydrodynamic equations, which express three conservation laws, namely of particle number (or the mass), momentum, and energy, and are valid over large space and time scales. This theory relies on the assumption of local thermal equilibrium: the system is divided into small cells, which are assumed to thermalise sufficiently fast with their environment.  However, in isolated \emph{integrable and near-intergrable} quantum systems -- wherein the 1D Bose gas is an example -- this assumption of fast thermalisation (i.e., relaxation to the conventional Gibbs ensemble of statistical mechanics) is not justified; instead, the system is expected to relax to the generalized Gibbs ensemble \cite{Rigol2007,rigol2008thermalization, experimental-GGE}, which respects the infinitely many conserved quantities of an integrable system, and not just three.
Moreover, simulations of the Euler equations of conventional hydrodynamics are often plagued with the gradient catastrophe problem, preventing them from adequately describing scenarios involving, for example, the formation of dispersive quantum shock waves  \cite{Doyon2017Large,Simmons2020,Dubessy2021UniversalShock}.

In 2016, two papers sparked the development of a hydrodynamic theory specialised to integrable systems, known as the theory of generalised hydrodynamics (GHD) \cite{ghydro1, ghydro2}. In contrast to CHD, GHD does not rely on the assumption of local thermal equilibrium in the canonical Gibbs ensemble sense. Instead, systems are assumed to relax to equilibrium states, described by a generalised Gibbs ensemble (GGE) \footnote{The term ``generalised'' in GHD is used in the same sense as in generalised Gibbs ensemble.}. GHD avoids the gradient catastrophe problem typical of scenarios involving, e.g., shock waves, and thus, can describe systems very far from equilibrium \cite{Doyon2017Large}. In particular, GHD has recently been shown to reproduce the most striking effects observed in the quantum Newton's cradle experiment in a strongly interacting 1D Bose gas \cite{Doyon2017Large,Caux2019_GHD_Quantum_Newton_Cradle}, such as the undamped collisional oscillations and the lack of conventional thermalization even after hundreds of collisions. The applicability of GHD extends well beyond ultra-cold atomic physics. Indeed, GHD has proven to be a robust theory applicable in both classical and quantum models over large length and time scales \cite{ghydro1,ghydro2,Doyon2017Dynamics,Doyon2017Large,Bulchandani2018Bethe,Bastianello2018Generalized,watson2022benchmarks}.

Since its original formulation, GHD has been extended to account for various experimentally relevant physical effects, such as weak integrability breaking in inhomogeneous systems \cite{Bastianello2019Inhomogeneous}, hydrodynamic diffusion \cite{DeNardis2018Diffusion,DeNardis2019Diffusion}, and quantum fluctuations \cite{Ruggiero2020QGHD}. This review aims to overview the theory of generalised hydrodynamics and its recent developments, as well as its laboratory tests in ultracold atom experiments. In doing so, we restrict ourselves to the applications of GHD to describe the dynamics of the repulsive one-dimensional (1D) Bose gas described by the Lieb-Liniger model \cite{Lieb-Liniger-I,Lieb-Liniger-II}. Our intention here is to give a brief yet reasonably comprehensive overview of this topical research area, intended for a broad, non-specialist audience. Upon doing so, we bring the attention of interested readers to the Special Issue \cite{bastianello2022introduction} of the Journal of Statistical Physics, devoted to recent advances in GHD. The Special Issue includes (among many original articles) several comprehensive, in-depth review articles on GHD intended for specialists \cite{Bastianello2021Hydrodynamics,alba2021generalized,Bulchandani2021Superdiffusion}; we also highlight a recent related review article by Guan et al. \cite{Guan2022New}, which includes a very brief overview of GHD in the broader context of new trends in quantum integrability. We hope that our review will occupy the space between these two extremes and that it will stimulate further interest in GHD in a broader physics community.

The organisation of this review is as follows. In section II, we introduce the Lieb-Liniger model and the equations of GHD at the Euler scale. We also describe the quantum Newton's cradle experiment, its description via GHD, and the implications on the mechanisms of thermalisation. In section III, we outline several extensions of the original formulation of GHD. In section IV, we review the experimental tests of GHD and the benchmarks against other theoretical approaches. Finally, section V offers some perspectives on the future direction of GHD and some open problems regarding GHD.

\section{Euler-scale Generalised Hydrodynamics}

\subsection{Theoretical considerations}

The Lieb-Liniger Hamiltonian for $N$ bosons in a uniform box of length $L$ (with periodic boundary condition) interacting via the two-body contact interaction potential $U(x,x') = g\delta(x-x')$ is given by \cite{Lieb-Liniger-I,*Lieb-Liniger-II}
\begin{equation}\label{eq:LL_Hamiltonian}
    H_{\text{LL}} = -\frac{\hbar^2}{2m}\sum_{i=1}^{N} \frac{\partial^2}{\partial x_i^2} + g\sum_{1\leq i<j<N}\delta(x_i-x_j),
\end{equation}
where $g$ quantifies the strength of interactions, assumed here to be repulsive ($g>0$). The Hamiltonian (\ref{eq:LL_Hamiltonian}) is integrable and exactly solvable using Bethe ansatz, and as such it admits an infinite number of conservation laws. In the second quantized form, it can be rewritten as
\begin{align}
	\hat{H}_{\text{LL}}	=&-\frac{\hbar^{2}}{2m}  \int dx\, \hat{\Psi}^{\dagger}(x) \frac{\partial^{2}}{\partial x^{2}} \hat{\Psi}(x)\nonumber\\
	&+ \frac{g}{2} \int dx\, \hat{\Psi}^{\dagger}(x) \hat{\Psi}^{\dagger}(x) \hat{\Psi}(x) \hat{\Psi}(x). \label{eq:H}
\end{align}

Such a one-dimensional system can be experimentally realised by confining an ultracold gas of bosons to, e.g., a highly elongated harmonic trap with transverse frequency $\omega_{\perp}$ and axial frequency $\omega \ll \omega_{\perp}$. When the transverse excitation energy is much larger than all other relevant energies of the problem, such as the average thermal energy and the chemical potential of the system, $\hbar\omega_{\perp}\gg \max\{k_BT,\, \mu$\}, the transverse excitations are negligible, and the dynamics take place only along the longitudinal dimension while being frozen out in the transverse dimension. For sufficiently large systems, the boundary effects can be neglected, so that the above Hamiltonian can describe the properties of systems that are not necessarily periodic or even uniform, wherein the inhomogeneities due to the longitudinal trapping $V(x)$ can be accounted for within the local density approximation (LDA) \cite{Kheruntsyan2005}.

An important parameter encoding the strength of interactions between particles in a uniform 1D Bose gas is the dimensionless Lieb-Liniger parameter $\gamma$ defined as 
\begin{equation}\label{eq:gamma_definition}
    \gamma = \frac{mg}{\hbar^2 n},
\end{equation} 
where $n=N/L$ is the 1D (linear) density. When $\gamma \ll 1$, the system is weakly interacting. Conversely, for $\gamma \gg 1$, the interaction energy is large, and the system is strongly interacting. Note that one can equivalently enter the strongly interacting regime $\gamma\gg 1$ by either increasing the interaction strength $g$ or decreasing the density of the system $n$.

In addition to the dimensionless interaction strength $\gamma$, one can also define a dimensionless temperature parameter, $\mathcal{T}$, by scaling the temperature of the system $T$ by the temperature of quantum degeneracy $T_d = \hbar^2 n^2/2mk_B $, 
\begin{equation}\label{eq:tau_definition}
    \mathcal{T} = \frac{T}{T_d} = \frac{2mk_BT}{\hbar^2 n^2}.
\end{equation}
When $T\sim T_d$, the thermal de Broglie wavelength of the particles is on the order of the mean interparticle separation. This represents the temperature regime below which quantum effects begin to dominate. Overall, finite temperature equilibrium properties of 1D Bose gas systems can be studied using the thermodynamics Bethe ansatz first derived and solved by Yang and Yang \cite{yang1969thermodynamics}.

To model the dynamics of a 1D Bose gas in the hydrodynamic sense, we view the system as a continuum of mesoscopic fluid cells that are thermodynamically large but small enough to be considered spatially homogeneous. For equilibrium states, this assumption is equivalent to the LDA. Due to integrability, our system admits a set of conserved charges $Q_i$ ($i=1,2,3,...$), such as energy, particle number, and momentum, each of which we assume can be written as an integral of a corresponding charge density, $q_i(x,t)$, i.e.,
\begin{equation}
    Q_i(t) = \int dx\, q_i(x,t).
\end{equation}
To each charge density, $q_i(x,t)$, we have a corresponding current density $j_i(x,t)$ satisfying a continuity equation,
\begin{equation}\label{eq:cont}
    \partial_t q_i + \partial_x j_i = 0.
\end{equation}
The fundamental assumption of GHD is that after some relaxation time, an inhomogeneous non-stationary system approaches, within each fluid cell, states which have maximised entropy with respect to each of the conserved quantities. These maximal entropy states are described by generalised Gibbs ensembles (GGE) with density matrices of the form, 
\begin{equation}
    \rho_{\text{GGE}}(x,t) \propto  e^{-\sum_{i}\beta^i(x,t)Q_i},
\end{equation}
where $\beta^i(x,t)$ is the Lagrange multiplier corresponding to the charge $Q_i$. The density matrix $\rho_{\text{GGE}}$ is related to the average of the charge density $q_i(x,t)$ via
\begin{equation}
    \mathfrak{q}_i(x,t) := \braket{q_i(x,t)} = \frac{\text{tr}\left(\rho_{\text{GGE}}\, q_i(x,t)\right)}{\text{tr}(\rho_{\text{GGE}})}.
\end{equation}
The set of average conserved densities $\{\mathfrak{q}_i\}_{i=1}^{\infty}$ can be considered as a set of coordinates for the manifold of maximal entropy states. In principle, this gives a complete (coarse-grained) description of our system as the set of average conserved densities specifies a particular configuration of our system, described as a point in the manifold of maximal entropy states. An alternative set of coordinates can be obtained by utilising the quasi-particle formulation of the thermodynamic Bethe ansatz (TBA) \cite{van2016introduction, yang1969thermodynamics}, in which the maximal entropy states are described via a phase space density of quasi-particles. As we will see, these coordinates offer a significant improvement over the set $\{\mathfrak{q}_i\}$ as the density of quasi-particles satisfies a single partial differential equation and once obtained, it is possible to construct each $\mathfrak{q}_i$. We note that whilst the techniques of the thermodynamic Bethe ansatz enable a description of the equilibrium properties of a uniform system, the applicability of GHD extends beyond this to non-equilibrium scenarios in spatially inhomogeneous systems within the so-called local density approximation (LDA), which is implicitly assumed in any hydrodynamic theory.

The main quantity of GHD is the density of quasi-particles, also known as the rapidity distribution, denoted $\rho_{p}(\theta,x,t)$ \footnote{In this review, we only consider many-body models of identical particles of single species, and hence, a single quasi-particle distribution. It is straightforward to generalise the equations of GHD to many species.}, that carry quasi-momentum $m\theta$ at the space-time point $(x,t)$. This quantity is analogous to the equilibrium root density in the thermodynamic Bethe ansatz solution. However, in GHD the rapidity distribution in our out-of-equilibrium problem evolves in time according to a classical Euler-like hydrodynamic equation. More specifically, the evolution of the rapidity distribution in GHD is governed by the set of integro-differential equations \cite{ghydro1, ghydro2, Doyon2017Note},
\begin{equation}\label{eq:GHD1}
\partial_t\rho_p + \partial_x\left(v^{\text{eff}}(\theta)\rho_p\right) - \frac1{m}\left(\partial_x V\right)\partial_{\theta}\rho_p= 0,    
\end{equation}
where $V(x)$ is an external potential, $v^{\text{eff}}(\theta)$ is the effective velocity, defined via the integral equation,
\begin{equation}\label{eq:GHD2}
    v^{\text{eff}}(\theta) = \theta + \int_{-\infty}^{\infty} d\theta'\, \Phi(\theta-\theta')\left( v^{\text{eff}}(\theta')- v^{\text{eff}}(\theta)\right)\rho_p(\theta'),
\end{equation}
and $\Phi(\theta - \theta')$ is the two-body scattering shift, which in the Lieb-Liniger model takes the form, 
\begin{equation}
\Phi(\theta) = \frac{\hbar^2}{m} \frac{2g}{g^2 + \hbar^2\theta^2}.  \end{equation}
Note we have suppressed the $x$ and $t$ dependence of $\rho_p$ and $v^{\text{eff}}$ above for simplicity. These are the main equations of GHD at the Euler scale, i.e., when quantities vary very slowly over space and time.

\begin{figure*}[t]
    \centering
    \includegraphics[width=17cm]{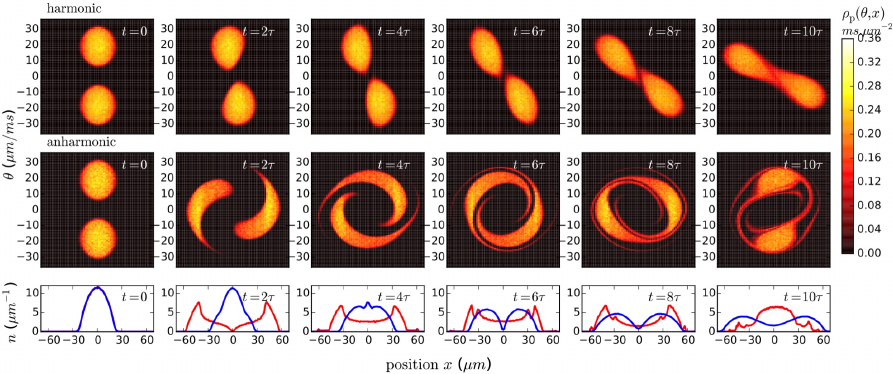}
    \caption{Evolution of the rapidity distribution over the first ten oscillation cycles for a strongly interacting 1D Bose gas in the quantum Newton's cradle scenario in a harmonic potential (top row) and with a small anharmonicity (middle row). Despite the presence of an integrability-breaking trapping potential and dephasing effects, neither system thermalises; the dephased states cannot be identified with a thermal state. The bottom row shows the corresponding density profiles, obtained by integrating the rapidity distribution $\rho_p(\theta, x)$ over all rapidities $\theta$. The blue curve corresponds to the harmonic potential, whilst the red curve is for the anharmonic trap. Adapted from \cite{Caux2019_GHD_Quantum_Newton_Cradle}}
    \label{fig:qNc-1}
\end{figure*}

Note that the effective velocity is a functional of the rapidity distribution and so two equations are coupled. Physically, this effective velocity can be interpreted as the large-scale, coarse-grained velocity of a quasi-particle as it travels through the gas, taking into account the scattering shifts it accumulates at collisions with the other quasi-particles \cite{Doyon-lectures}. For the uniform Lieb-Liniger gas, the initial rapidity distribution $\rho_p(\theta, x, 0)$ that is supplied to the GHD equations is often taken as the thermal equilibrium distribution obtained via the Yang-Yang thermodynamic Bethe ansatz \cite{yang1969thermodynamics}. For the nonuniform case, the initial rapidity distribution can be found locally for each $x$ using the same thermodynamic Bethe ansatz but within the LDA \cite{Kheruntsyan2005}. 

We note that although GHD was first introduced for quantum field theories \cite{ghydro1} and quantum chains \cite{ghydro2}, equations (\ref{eq:GHD1}) and (\ref{eq:GHD2}) have previously been derived rigorously in the context of the classical hard rod gas \cite{percus1976equilibrium, boldrighini1983one} and in soliton gases \cite{El2005Soliton,EL2003374,Doyon2018Soliton}. 

Given the rapidity distribution, $\rho_p(\theta, x,t)$, at some fixed time, the expectation value of the conserved charge densities and their corresponding currents can then be computed via 
\begin{align}
    \mathfrak{q}_i(x,t) &= \int d\theta\, h_i(\theta)\,  \rho_{p}(\theta), \label{eq:old_charge}\\
    \mathfrak{j}_i(x,t) &= \int d\theta\, h_i(\theta)\, v^{\text{eff}}(\theta)\, \rho_{p}(\theta). \label{eq:old_current}
\end{align}
where $h_i(\theta)$ is the one-particle eigenvalue of the $i$-th conserved charge. For example, the average particle density can be obtained by setting $h_i(\theta) = 1$, whilst the average energy density is found by letting $h_i(\theta) = m\theta^2/2$. Intuitively, equations (\ref{eq:old_charge}) and (\ref{eq:old_current}) can be understood as follows: the contribution of a  quasi-particle of rapidity $\theta$ to the $i$-th conserved charge is $h_i(\theta)$, moreover, the number of quasi-particles with rapidity in the interval $[\theta,\theta+d\theta)$ is given by $\rho_p(\theta)d\theta$. Therefore, the contribution of all quasi-particles with rapidity in the range $[\theta, \theta+d\theta)$ to the $i$-th charge is given by $h_{i}(\theta)\rho_{p}(\theta)d\theta$. Summing over all such rapidities yields Eq.~(\ref{eq:old_charge}). A similar interpretation holds for Eq.~(\ref{eq:old_current}) for the average current density. We note also that the evolution equation for the rapidity distribution in Eq.~(\ref{eq:GHD1}) follows immediately by substituting Eq.~(\ref{eq:old_charge}) and Eq.~(\ref{eq:old_current}) into the respective continuity equations and using completeness of the set of functions $\{h_i(\theta)\}$.

Although Eq.~\eqref{eq:old_current} was proposed in the original papers that introduced the theory of GHD \cite{ghydro1,ghydro2}, its numerical verification \cite{Doyon2017Dynamics,Bastianello2018Generalized,Bastianello2019Inhomogeneous}, as well as rigorous derivations and proofs of $v^{\text{eff}}(\theta)$ as the equation of state of GHD \cite{Yoshimura2019EoS,Drude_weight_2019,spohn2020collision,Pozsgay_PRL_2020,Pozsgay_PRX_2020,Pozsgay_SciPost_2020}, were provided later. For a recent review of current operators of one-dimensional integrable models, we direct the reader to the review \cite{borsi2021current}.

It is often more convenient to encode the thermodynamic properties of a system via the filling factor, 
\begin{equation}
    \vartheta(\theta,x,t) := \frac{\rho_{p}(\theta)}{\rho_p(\theta) + \rho_h(\theta)} = \frac{\rho_p(\theta)}{\rho_s(\theta)}.
\end{equation}
where $\rho_h$ is the density of `holes' and $\rho_s(\theta)$ is the density of states which is related to the density of quasi-particles, $\rho_p$, via the thermodynamic Bethe equation,
\begin{equation}
    2\pi \rho_s(\theta) = 1 + \int_{-\infty}^{\infty}d\theta'\, \Phi(\theta - \theta')\rho_{p}(\theta'). 
\end{equation}

One can then construct the average charge and current densities using
\begin{align}
    \mathfrak{q}_i(x,t) &= \int \frac{d\theta}{2\pi}\, \vartheta(\theta)\, h^{\text{dr}}_i(\theta),\\
    \mathfrak{j}_i(x,t) &= \int \frac{d\theta}{2\pi}\,  \vartheta(\theta)\, \theta\, h^{\text{dr}}_i(\theta), 
\end{align}
where the dressing operation $f\mapsto f^{\text{dr}}$ is defined via the integral equation, 
\begin{equation}
    f^{\text{dr}}(\theta) = f(\theta) + \int\frac{d\theta'}{2\pi}\, \Phi(\theta' - \theta)\vartheta(\theta')f^{\text{dr}}(\theta').
\end{equation}
In this formulation, Eq.~(\ref{eq:GHD1}) becomes,
\begin{equation}\label{eq:GHD_filling}
\partial_t \vartheta + v^{\text{eff}} \partial_x \vartheta - \frac1{m}\left(\partial_x V\right)\partial_{\theta}\vartheta= 0,    
\end{equation}
where the effective velocity is now outside the spatial derivative. Numerically, Eq.~(\ref{eq:GHD_filling}) is more convenient to work with than Eq.~(\ref{eq:GHD1}). We also advertise the open-source Matlab framework ``iFluid" that numerically solves the equations of GHD \cite{ifluid}.

\subsection{The quantum Newton's cradle experiment}

Arguably, the lack of thermalisation in isolated quantum systems was best demonstrated in the hallmark quantum Newton's cradle experiment of Kinoshita 
\textit{et al.} \cite{kinoshita2006quantum}. There, clouds of strongly interacting rubidium atoms confined to a one-dimensional harmonic trap undergo repeated collisions without noticeably thermalising on observable time scales -- corresponding to thousands of collisions. The lack of thermalisation can be attributed to the integrability of the underlying Lieb-Liniger model in the uniform limit and to weak integrability in the nonuniform (harmonically trapped) system owing to the applicability of the local density approximation \cite{Kheruntsyan2005}.

In the decade that followed the Kinoshita experiment, a quantitatively accurate model of the quantum Newton's cradle with experimentally relevant parameters remained elusive. GHD has emerged as an ideal tool to model this experiment. In 2018, Caux \textit{et al.} used GHD to simulate the dynamics of a strongly interacting 1D Bose gas in the quantum Newton's cradle experiment \cite{Caux2019_GHD_Quantum_Newton_Cradle}. The evolution of the rapidity distribution is shown in Fig.~\ref{fig:qNc-1} for the case of a harmonic potential and an anharmonic potential to mimic the trapping potential in the original quantum Newton's cradle experiment. 

Caux \textit{et al.} found that even in the presence of a trapping potential that weakly breaks integrability, the 1D gas does not thermalise. Instead, the system relaxes to a generalised Gibbs ensemble. The preclusion of thermalisation here can be attributed to the existence of conserved quantities that are incompatible with convergence towards thermal equilibrium \cite{Caux2019_GHD_Quantum_Newton_Cradle}. These quantities take the form
\begin{equation}\label{eq:cons_quantity}
 Q[f] = \int d\theta\, dx\, f(\vartheta(\theta, x, t)) \rho_p(\theta, x,t),   
\end{equation}
where $f$ is an arbitrary function and $\rho_p(\theta, x, t)$ is continuous in $\theta$ and $x$. The fact that these quantities are conserved under GHD evolution in a trap follows directly from Eq.~(\ref{eq:GHD1}) and Eq.~(\ref{eq:GHD_filling}). We note that these quantities are only conserved at the Euler scale.

\section{Beyond Euler-scale 
Generalised Hydrodynamics}
Here we present some of the extensions to the original formulation of GHD. Namely, the incorporation of diffusive and quantum effects, the description of dimensional crossover in one-dimensional Bose gases, and the inclusion of space-time dependent interactions. 

\subsection{Diffusive effects}

Despite the excellent predictive power of standard GHD, it is often necessary to go beyond the lowest-order Euler scale. For instance, spin and charge transport in quantum chains have been shown to exhibit diffusion, and other non-Eulerian behaviours not captured by Euler scale GHD \cite{PhysRevE.96.020105, PhysRevLett.103.216602, ljubotina2017spin}.

\begin{figure}[h!]
    \centering
    \includegraphics[width=6.3cm]{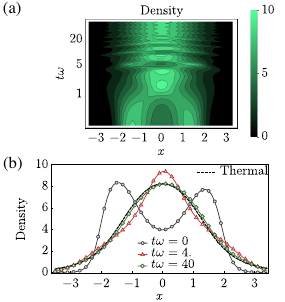}
    \caption{Simulation of diffusive GHD for a weakly interacting 1D Bose gas in the quantum Newton's cradle setup initiated via a quench of a double-well trap to a single harmonic well. (a) Evolution of the density profile. (b) The density profile at three fixed times compared to its distribution in thermal equilibrium. In contrast to Euler-scale GHD, this system eventually thermalizes over sufficiently large time scales when diffusive effects are considered. Adapted from \cite{Bastianello2020Thermalisation}.}
    \label{fig:diffusion-GHD}
\end{figure}

As a first approximation, we assume that the state of the system and local observables at the point $(x,t)$ can be described by the averages of conserved charges in a neighbourhood of this point. We are then permitted to take a gradient expansion of the local observables around $(x,t)$. In particular, averages of the current densities, $\mathfrak{j}_i(x,t)$ can be written in the form,
\begin{equation}\label{eq:gradient-expansion}
    \mathfrak{j}_i(x,t) = \mathcal{F}_{i}(x,t) + \sum_{j}\mathcal{F}_{ij}(x,t)\partial_x \mathfrak{q}_j(x,t) + \dots,
\end{equation}
where the space-time dependence of the functions $\mathcal{F}_i$, $\mathcal{F}_{ij}$, etc. is encoded through the charge densities $\mathfrak{q}_i(x,t)$. Retaining only the first order term $\mathcal{F}_i$ corresponds to the so-called Euler-scale GHD. The next higher-order correction is the diffusive Navier-Stokes correction. This term introduces an arrow of time through the (irreversible) production of entropy. The corresponding Navier-Stokes diffusive GHD equation was first derived by De Nardis \textit{et al.} \cite{DeNardis2018Diffusion} (see also \cite{DeNardis2019Diffusion,Bastianello2020Thermalisation,Durnin2021Diffusive}), and reads
\begin{equation}\label{eq:diffusion-GHD}
    \partial_t \rho_p + \partial_x\left(v^{\text{eff}}\rho_p\right) = \partial_x\left(\mathfrak{D}\partial_x\rho_{p}\right) + \frac1{m}\partial_x V\partial_\theta \rho_p,
\end{equation}
where $\mathfrak{D}$ denotes the integral operator defined by the action, 
\begin{equation}
    \mathfrak{D}f(\theta) = \int d\theta' \, \mathcal{D}(\theta, \theta') f(\theta'),
\end{equation}
and $\mathcal{D}(\theta,\theta')$ is the diffusion kernel which satisfies the relation, 
\begin{equation}
    [\mathcal{D}(\theta, \cdot)]^{\text{dr}}(\theta')\rho_s(\theta') = [\rho_s(\cdot)\tilde{\mathcal{D}}(\cdot, \theta')]^{\text{dr}}(\theta),
\end{equation}
with 
\begin{align}
    &\tilde{\mathcal{D}}(\theta, \theta') = \delta(\theta-\theta')\left[ \int d\alpha\, \rho_p(\alpha)(1-\vartheta(\alpha))\right. \notag  \\
    &\,\,\,\,\,\,\quad\quad\quad\quad \times \left. \left(\frac{\Phi^{\text{dr}}(\alpha-\theta)}{\rho_s(\theta)}\right)^2\left|v^{\text{eff}}(\alpha) - v^{\text{eff}}(\theta) \right|\right]\notag \\
    &\quad - \rho_p(\theta)(1-\vartheta(\theta))\left(\frac{\Phi^{\text{dr}}(\theta-\theta')}{\rho_s(\theta)}\right)^2\left|v^{\text{eff}}(\theta) - v^{\text{eff}}(\theta') \right|,
\end{align}
which we note are all quantities from the Euler-scale GHD \cite{DeNardis2018Diffusion}. Physically, the diffusion kernel arises due to two-body scattering processes among quasi-particles \cite{DeNardis2019Diffusion} that are neglected in the Euler-scale hydrodynamics. Such two-body scattering processes, however, become important at smaller length scales, leading to the decay of current-current correlations and therefore to the presence of finite diffusion constants.

\begin{figure}[tp]
    \centering
    \includegraphics[width=7.2cm]{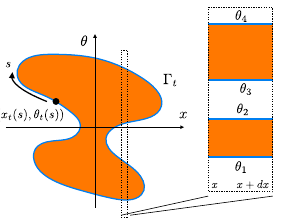}
    \caption{Fermi contour of a system at zero temperature. The orange region $\Gamma_t$ corresponds to a unit filling factor $\vartheta(x,\theta) = 1$. The dynamics of the system are reduced to that of the contour $\partial \Gamma_t$. Locally, the region $\Gamma_t$ is split into disjoint Fermi seas defined by the set of Fermi points $\{\theta_1,\dots, \theta_{2n}\}$. Adapted from \cite{Ruggiero2020QGHD}.}
    \label{fig:fermi-contour}
\end{figure}

Using diffusive GHD, Bastianello \textit{et al.} \cite{Bastianello2020Thermalisation} studied the thermalisation of a 1D Bose gas in the quantum Newton's cradle experiment, see Fig.~\ref{fig:diffusion-GHD}. The authors found that diffusion was the leading mechanism that induced thermalisation towards a stationary state in the presence of an (integrability-breaking) trapping potential. The system initially relaxes to a pre-thermal state described by a GGE, and then at much longer, diffusive time scales (which scale with the length of the system $L$ as $\propto L^2$) it eventually relaxes to the thermal state described by the standard Gibbs ensemble. These results demonstrate the fundamental role that diffusive effects play in the late-time dynamics of near-integrable systems.

\begin{figure*}[t]
    \centering
    \includegraphics[width=14.5cm]{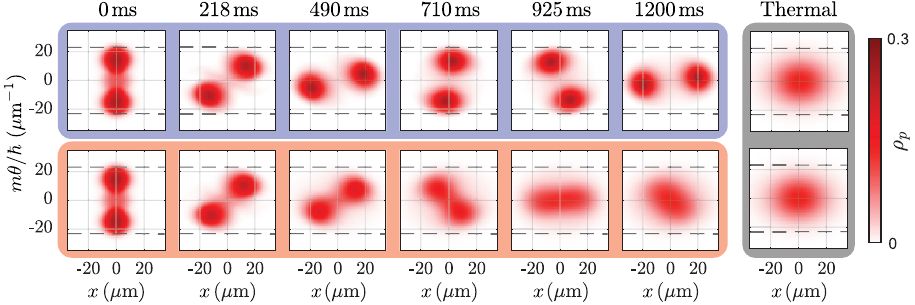}
    \caption{Evolution of the rapidity distribution for a 1D Bose gas containing 130 atoms at $94$ nK in a quantum Newton's cradle setup initiated with Bragg pulses. The top row shows the evolution during the first 100 oscillation periods according to the equations of standard GHD. The bottom row shows the evolution according to the extended model of M\o{}ller \textit{et al} that incorporates population in transversely excited states and hence leads to faster thermalization. The dashed lines mark the excitation threshold. The small fraction of transverse excited atoms has a strong influence on the dynamics of the system; the inclusion of the collision integral enables quasi-particles to re-distribute across phase space and thermalize. The last two columns show the best fit for a thermal state at the temperature of the final evolved system. Adapted from \cite{Moller2021Extension}. }
    \label{fig:dimensional-crossover-results}
\end{figure*}

\subsection{Quantum Generalised Hydrodynamics}

 In its original formulation, GHD neglects important quantum effects, such as quantum fluctuations and entanglement entropy. A fundamental assumption of GHD is that each fluid cell comprising the system is independent at any fixed time. This implies that equal-time correlations between fluid cells vanish. However, as with many quantum systems, these equal-time correlations are typically non-zero \cite{RevModPhys.83.1405}; such effects occur beyond the Euler scale. One may resolve these shortcomings by re-quantising the theory of GHD. This was first achieved by Ruggiero \textit{et al.} in 2020 in Ref.~\cite{Ruggiero2020QGHD}, in which they considered a system initially at zero temperature, for which quantum fluctuations are most significant. Since the entropy is initially zero and is conserved under the evolution of the standard GHD equations, the state remains at zero entropy at all times.

Zero-entropy states are characterised by the so-called Fermi contour $\partial \Gamma_t$ enclosing the region $\Gamma_t$ in $(x,\theta)$ phase space with unit filling, 
\begin{equation}
    \vartheta(x, \theta, t) = \begin{cases}
    1, & \text{if } (x,\theta)\in \Gamma_t,\\
    0, & \text{otherwise}.
    \end{cases}
\end{equation}
Locally, a zero-entropy state can be described by a split Fermi sea, consisting of a collection of disjoint regions in $(x,\theta)$ phase space for which the filling factor is unity, 
\begin{equation}
    \vartheta(x,\theta, t) = \begin{cases}
    1,& \text{if } \theta \in [\theta_1,\theta_2]\cup \dots \cup [\theta_{2p-1}, \theta_{2p}]\\
    0,&\text{otherwise},
    \end{cases}
\end{equation}
where $\theta_{n} \equiv \theta_n(x,t)$, $n\in \{1,2,\dots, 2p\}$ denote the Fermi points at position $x$ and time $t$,
see Fig.~\ref{fig:fermi-contour}.

For these zero-entropy states, it was shown by Doyon \textit{et al.} that the equations of GHD, Eq.~(\ref{eq:GHD_filling}), reduce to a set of equations for the Fermi points \cite{Doyon2017Large},
\begin{equation}
\partial_t \theta_n + v^{\text{eff}}\partial_x\theta_n = -\frac1{m}\partial_x V.
\end{equation}
One then considers small fluctuations about the Fermi contour. Locally, the Fermi points are modified via $\theta_n \to \theta_n + \delta \theta_n$, where $\delta\theta_n$ satisfies the differential equation, 
\begin{equation}
    (\partial_t + v^{\text{eff}}\partial_x)\delta\theta_n(x,t) = 0,
\end{equation}
describing the propagation of linear sound waves \cite{ruggiero2021quantum}.

To re-quantise this theory, one promotes the fluctuations about the Fermi contour to linear operators acting on a Hilbert space, i.e., $\delta\theta_n \to \delta\hat{\theta}_n$. The key insight of Ruggiero \textit{et al.} is that the problem of quantising sound waves above the classical GHD ground state can be recast as a problem of quantising incompressible regions in phase space, which is well-known in the literature on the quantum Hall effect \cite{PhysRevB.41.12838}. The resulting re-quantised theory is a time-dependent, spatially inhomogeneous, multi-component Luttinger liquid, and has been dubbed ``quantum generalised hydrodynamics''. It describes quantum fluctuations of non-equilibrium systems where conventional Luttinger liquid theory fails \cite{Ruggiero2020QGHD}. Additionally, unlike the Luttinger liquid theory, this quantum GHD is not restricted to low energies, and it also is no longer restricted to zero entropy states \cite{ruggiero2021quantum,Vlijm2016Colleralions}.

\subsection{Dimensional crossover for the 1D Bose Gas}
\label{sec:Dimensional_crossover}

One-dimensional atomic gases are often realised by using highly elongated ``cigar-shaped" trapping potentials. It is assumed that due to the strong radial confinement, the transverse modes are effectively frozen out. However, when two atoms of sufficiently large momenta collide, the collisional energy may exceed the level spacing of the transverse confinement. This leads to a non-zero population of transverse excited states, which breaks integrability. To account for collisions of atoms in transverse excited states, M{\o}ller \textit{et al.} introduced a phenomenological Boltzmann-type collision integral into the equations of GHD \cite{Moller2021Extension}.

M{\o}ller \textit{et al.} studied the dynamics of a quasi-1D Bose gas in the quantum Newton’s cradle scenario using this modified GHD. The evolution of the rapidity distributions under the equations of standard  GHD and the extended GHD are shown in Fig.~\ref{fig:dimensional-crossover-results}. By accounting for collisions with atoms in transverse excited states, the modified GHD predicts that the system thermalizes. This is in contrast to standard GHD which does not predict thermalization. The rate of thermalization of this extended GHD was also found to be consistent with previous experimental observations \cite{10.21468/SciPostPhys.9.4.058}.

A further test of GHD in the quasi-1D regime was provided by Cataldini \text{et al.}  \cite{Cataldini2022Emergent} which demonstrated that GHD can accurately describe the dynamics of a Bose gas whose chemical potential and thermal energy far exceed the conventional limits of one-dimensionality. In this experiment, a weakly-interacting quasi-1D Bose gas, initially confined to a 1D box trap with a sinusoidal bottom, was suddenly quenched to a flat bottom trap. The evolution of the density profile was then measured and showed excellent agreement with the predictions of standard GHD. The observed agreement was attributed to the occupation of low rapidity states which, by virtue of the fermionic quasi-particle statistics, reduces the number of collision channels for transverse excitations, resulting in an emergent Pauli blocking of transverse excitations. The predictions of standard GHD and the extended GHD in this setup were almost identical. This is in contrast to the previously mentioned experiment of M{\o}ller \textit{et al.} \cite{Moller2021Extension}. The difference here arises due to the initial double-peaked rapidity distribution in the quantum Newton's cradle setup which permits virtual transitions to unoccupied transverse excited states, not blocked by the effective Pauli exclusion.

\subsection{Space-time inhomogeneous interactions}

Inhomogeneities are ubiquitous in experiments. The effects of a (spatially-varying) trapping potential were first introduced into GHD by Doyon and Yoshimura in 2017 \cite{Doyon2017Note} in which the authors considered the addition of a generalised potential to the Hamiltonian, 
\begin{equation}
    H \to H + \sum_{k} \int dx\, V_k(x) q_k(x).
\end{equation}
This includes the effects of, e.g., the standard external trapping potential $V(x)$ for $k=0$, or perturbations by an inhomogeneous temperature field, by including the energy density $q_1(x)$. However, there exists another relevant inhomogeneity not addressed in standard GHD, namely, that of space-time inhomogeneous interactions. Bastianello \textit{et al.} were the first to incorporate these effects into GHD. They found that these inhomogeneities introduce an additional term into the standard GHD equations \cite{Bastianello2019Inhomogeneous}, 
\begin{align}
    \partial_t\rho_p + \partial_x\left(v^{\text{eff}}\rho_p\right) + \partial_{\theta}\left(\frac{f^{\text{dr}}\partial_t\alpha + \Lambda^{\text{dr}}\partial_x\alpha}{(\partial_{\theta}p)^{\text{dr}}}\right) \nonumber\\
    = \frac1{m}\left(\partial_x V\right)\partial_{\theta}\rho_p,
\end{align}
where $\alpha(x,t)$ is any parameter in the Hamiltonian that can be varied whilst still maintaining integrability (for instance, the interaction strength $g$ in the Lieb-Liniger model), whereas the forces $f$ and $\Lambda$ are given by,
\begin{align}
    f(\theta) &= - \frac{\partial p(\theta)}{\partial \alpha} + \int d\theta'\, \frac{\partial \Phi(\theta-\theta')}{\partial \alpha} \rho_p(\theta'),\label{eq:inhom_1}\\
    \Lambda(\theta) &= - \frac{\partial \varepsilon(\theta)}{\partial \alpha}   
     +  \int d\theta'\, \frac{\partial \Phi(\theta-\theta')}{\partial \alpha} v^{\text{eff}}(\theta')\rho_p(\theta'),\label{eq:inhom_2}
\end{align}
where $\varepsilon(\theta)$ and $p(\theta)$ are the energy and momentum eigenvalues, respectively.
We note that the above equations hold for general integrable models. For the Lieb-Liniger model, in particular, one has $p(\theta) = m\theta$ and $\varepsilon(\theta) = \frac12 m\theta^2 - \mu $, hence the first terms in Eqs.~(\ref{eq:inhom_1}) and (\ref{eq:inhom_2}) vanish. However, such inhomogeneities are present in, for example, the classical sinh-Gordon model. These results exhaust all possible inhomogeneities which
can be considered on a purely hydrodynamical level \cite{Bastianello2019Inhomogeneous}.

 \begin{figure}[tbp]
    \centering
    \includegraphics[width=6.5cm]{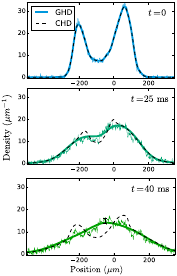}
    \caption{Measurements of the \textit{in situ} density profile for a system of $N=6300\pm 200$ bosons, initially trapped in a 1D double-well potential and subsequently released to freely expand in 1D. These results (noisy lines) are compared against the predictions of conventional (dashed lines) and generalised hydrodynamics (solid lines). Conventional hydrodynamics incorrectly predicts two distinct density peaks at late times. By contrast, GHD shows excellent agreement with the experimental data. Adapted from \cite{GHD_Atom_Chip}. }
    \label{fig:GHD_atom_chip}
\end{figure}

 \begin{figure*}[tbp]
    \centering
    \includegraphics[width=13.8cm]{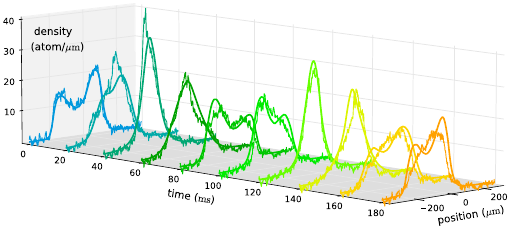}
    \caption{Measurements of the \textit{in situ} density profile for a system of $N=3500\pm 140$ atoms, quenched from a 1D double-well potential to a harmonic potential \cite{GHD_Atom_Chip}. These results are compared against the predictions of GHD shown as smooth lines on top of noisy experimental data. While the standard GHD reproduces most of the main features of the emergent dynamics, the minor disagreement with experimental data at later times can be rectified  \cite{Moller2021Extension} by accounting for transverse excitations of the gas using the 
    extension of GHD to dimensional crossover as 
    outlined in Section  \ref{sec:Dimensional_crossover}. Adapted from \cite{GHD_Atom_Chip}. }
    \label{fig:GHD_atom_chip2}
\end{figure*}

\section{Experimental Tests and Benchmarks of Generalised Hydrodynamics}
 
To date, there have been four major experimental tests of GHD \cite{GHD_Atom_Chip,Malvania2021GHD,Moller2021Extension,Cataldini2022Emergent}. These have all been within the context of 1D Bose gases confined to highly anisotropic trapping potentials as to realise the Lieb-Liniger model. In this section, we outline two of these experiments \cite{GHD_Atom_Chip,Malvania2021GHD} and their descriptions via GHD, while the other two experiments \cite{Moller2021Extension,Cataldini2022Emergent} have already been mentioned in Section \ref{sec:Dimensional_crossover} in the context of dimensional crossover in quasi-1D Bose gases.

\subsection{Tests of GHD in the weakly interacting regime}

In 2019, Schemmer \textit{et al.} experimentally demonstrated the validity of GHD for a system of one-dimensional bosons realised on an atom chip
\cite{GHD_Atom_Chip}. There, approximately $6300\pm 200$ particles were confined to one dimension via magnetic trapping techniques developed by \cite{PhysRevLett.83.3398} and \cite{PhysRevLett.84.4749}. The particles were initially confined in a double-well potential at thermal equilibrium and were subsequently allowed to expand freely. The \textit{in situ} density profile is shown at three times during the evolution in Fig.~\ref{fig:GHD_atom_chip}. One finds that GHD provides an excellent description of the evolving density profile, whilst conventional hydrodynamics incorrectly predicts the formation of two sharp density peaks.

Schemmer \textit{et al.} also considered the dynamics of a system of 1D bosons induced via a quench from a double-well potential to a harmonic potential. They found that GHD reproduces most of the main features of the emergent dynamics, see Fig.~\ref{fig:GHD_atom_chip2}. However, GHD predicts that the two density peaks persist much longer than the experimental data. Schemmer \textit{et al.} attributed this discrepancy to atom losses in the experiment, such as three-body losses due to three-body recombination processes \cite{soding, PhysRevLett.92.190401}, that break integrability. It was found that the total number of atoms in their system was reduced by 15\% during the dynamics. It has been argued by M\o{}ller \textit{et al.} that transverse excitations may have also influenced the observed dynamics \cite{Moller2021Extension}. By using the extension of GHD to the dimensional crossover regime, as outlined in Section \ref{sec:Dimensional_crossover}, M\o{}ller \textit{et al}. found that the density peaks were less pronounced than in standard GHD, in agreement with the experimental data.

\subsection{Tests of GHD in the strongly interacting regime}

In the strongly interacting regime -- the opposite regime to that of the atom chip experiment of Schemmer \textit{et al.} -- Malvania \textit{et al.} used a  2D lattice of independent harmonically trapped 1D Bose gases to test the predictions of GHD for the rapidity distributions following a strong confinement quench
\cite{Malvania2021GHD}.

\begin{figure}[bh]
    \centering
    \includegraphics[width=8.5cm]{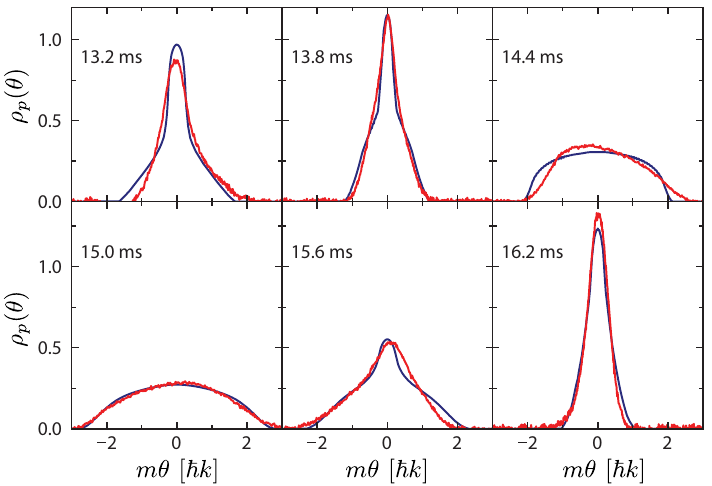}
    \caption{Comparison of experimental and GHD rapidity distributions of a strongly interacting 1D Bose gas following a strong confinement quench. The rapidity distributions are shown at different times, as indicated in each panel. The red curves show the experimental data, whilst the blue lines are the predictions of GHD, demonstrating good overall agreement between the theory and experiment. Adapted from \cite{Malvania2021GHD}. }
    \label{fig:GHD-experiment_strong}
\end{figure}

\begin{figure}[tbp]
    \centering
    \includegraphics[width=8.3cm]{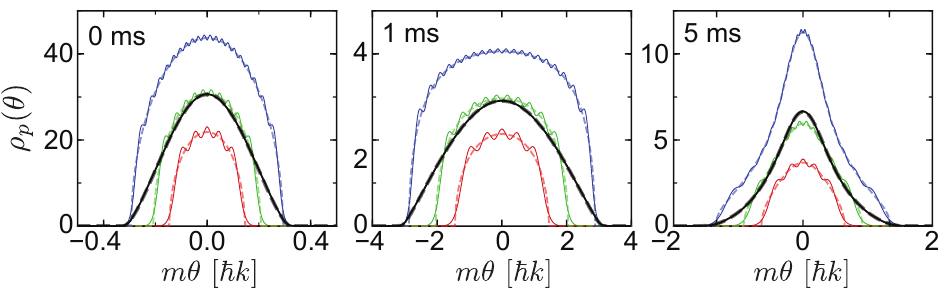}
    \caption{Comparison of the post-quench rapidity distributions from GHD (dashed lines) and from exact dynamics (solid lines) in the Tonks-Girardeau limit of infinite interaction strength for small atom numbers. The red, green, and blue lines correspond to 1D tubes with $5$, $10$, and $20$ atoms, respectively. The black lines are averages of the other lines. Adapted from \cite{Malvania2021GHD}.}
    \label{fig:GHD-experiment_2_N_low}
\end{figure}

In this experiment, bundles of one-dimensional gases of rubidium-87 atoms were generated using optical trapping potentials. The average number of atoms per gas was on the order of 10. This ensures that the particle number density $n$ is low and hence the system is well within the strongly interacting regime ($\gamma\gg 1$). The rapidity distribution was then measured by first switching off the axial trapping potential and letting the atoms freely expand purely in 1D until the rapidity distribution evolves into the momentum distribution \cite{SuddenExpansion,Campbell2015Sudden}. This is then followed by switching off the transverse confinement as well hence allowing for the measurement of the resulting momentum distribution via the subsequent time of flight expansion \cite{Wilson2020Dynamical}. The evolution of the space-integrated  rapidity distribution [$\rho_p(\theta,t)=\int dx \rho_p(\theta,x,t)$] measured in this way is shown in Fig.~\ref{fig:GHD-experiment_strong}. Note that the reported rapidity distribution is an average over all 1D tubes in the lattice. It was found that the dynamics of the gases were well-described by GHD. This is surprising as the low number of particles challenges the fundamental hydrodynamic assumption of GHD, which assumes that each fluid cell is thermodynamically large.

To address the unreasonable predictive power of GHD at low particle numbers, Malvania \textit{et al.} numerically simulated the experiment for a 1D Bose gas with a significantly stronger interaction strength, for which a comparison with exact results is possible. They found that the dynamics predicted by GHD were essentially  identical to that obtained exactly, up to small ripples (Friedel oscillations) in the rapidity distribution, which arise at low particle numbers, see Fig.~\ref{fig:GHD-experiment_2_N_low}.

\subsection{Benchmarks of generalised hydrodynamics}

In this section, we compare the predictions of GHD against established theoretical approaches in the context of out-of-equilibrium dynamics of 1D Bose gases. Through these comparisons, we aim to elucidate the strengths and limitations of GHD.

\begin{figure}[tbp]
    \centering
    \includegraphics[width=8.2cm]{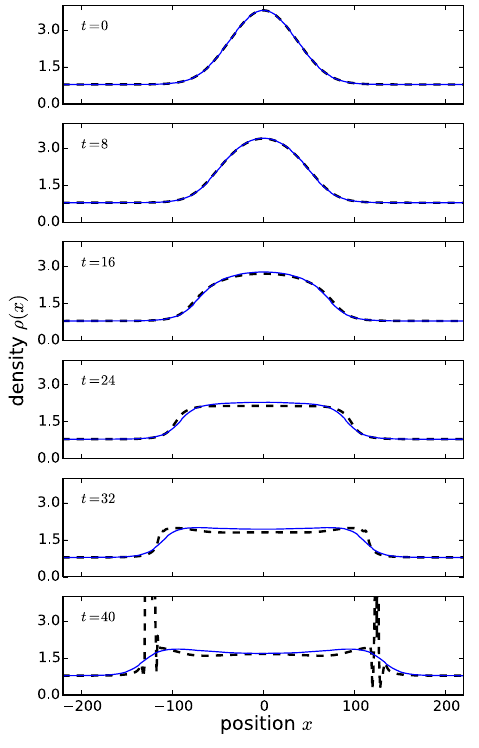}
    \caption{Free evolution of a localized density bump on a uniform background in a finite temperature 1D Bose gas. The predictions of CHD (black, dashed) are compared with those of GHD (blue). The position ($x$) and time ($t$) variables are in dimensionless units corresponding to $\hbar=m=1$, and the dimensionless temperature and interaction strength are chosen to be $T=1$ and $c=mg/\hbar^2=2$.  At large times, CHD develops two shocks (derivative discontinuities) on the two counter-propagating wave-fronts and is unable to describe subsequent dynamics. GHD remains smooth at all times and continues to evolve past the onset of the classical shock. Adapted from \cite{Doyon2017Large}.}
    \label{fig:CHD_comparison}
\end{figure}

An early comparison between GHD and CHD for the 1D Bose gas was provided by Doyon \textit{et. al} in Ref.~\cite{Doyon2017Large}, in which they demonstrated that GHD surpasses the theory of conventional (or classical) Euler hydrodynamics in its ability to describe the dynamics of dissolution of an initial localised density bump on a uniform background. In Fig.~\ref{fig:CHD_comparison}, we show the results of such a comparison, where we see that at early times the predictions of CHD and GHD are nearly identical. However, at longer times CHD develops a shock that manifests itself as a derivative discontinuity or the so-called gradient catastrophe problem, common to classical Euler hydrodynamics that
ignores dissipation and dispersion.
This implies that Euler CHD is unable to describe the dynamics of the systems past this time. The standard GHD profile, by contrast, remains smooth and continues to describe the dynamics of dissolution past the classical shock time, despite the fact that it also ignores dissipation and dispersion effects. The reason for the success of GHD compared to CHD is that it respects the conservation of infinitely many local charges, instead of just three charges conserved in CHD which are the particle number, momentum, and energy. By doing so, GHD avoids the development of the unphysical gradient catastrophe problem.

Related comparisons of GHD with CHD were done also in Refs.~\cite{doyon2018geometric,GHD_Atom_Chip} (see also Fig.~\ref{fig:GHD_atom_chip}) which demonstrated similar behaviours and arrived at the same conclusions. Other benchmarks of GHD for 1D Bose gases included comparisons with full quantum simulations at zero temperature using the numerical ABACUS algorithm \cite{Doyon2017Large}, the mean-field Gross-Pitaevskii (or nonlinear Schr\"{o}dinger) equation in the weakly interacting regime  \cite{GHD_Atom_Chip,Bastianello_dephasing_2020,Dubessy2021UniversalShock,watson2022benchmarks}, exact numerics in the Tonks-Girardeau regime of infinitely strong interaction strength \cite{Malvania2021GHD,Dubessy2021UniversalShock,watson2022benchmarks} (see also Fig.~\ref{fig:GHD-experiment_2_N_low}), and the matrix product state (MPS) or time-dependent density matrix renormalisation group ($t$DMRG) methods at intermediate interactions \cite{Ruggiero2020QGHD,ruggiero2021quantum,watson2022benchmarks}.

The initial benchmarks of GHD were in dynamical scenarios where generalised hydrodynamics was expected to be a valid theory. In all these scenarios GHD has indeed  demonstrated very good agreement with the alternative approaches. On the other hand, in scenarios involving small length-scale phenomena (which are not captured by GHD), it was conjectured that GHD would nevertheless adequately describe coarse-grained averages of the more accurate microscopic theories \cite{Doyon2017Large,GHD_Atom_Chip,Moller2021Extension}. A specific model of such coarse-grained averaging that mimics finite imaging resolution in ultracold atom experiments was proposed and analysed by Watson \textit{et al.} \cite{watson2022benchmarks}. Watson \textit{et al.} have also scrutinised the performance of GHD by benchmarking it against an array of alternative theoretical methods (including those applicable for finite temperature initial states) in some of the most challenging dynamical scenarios, such as the propagation of dispersive quantum shock waves \cite{Simmons2020,Dubessy2021UniversalShock} and collisional dynamics in various quantum Newton's cradle setups \cite{kinoshita2006quantum,GHD_Atom_Chip}.

\begin{figure}[tbp]
    \centering
    \includegraphics[width=8.2cm]{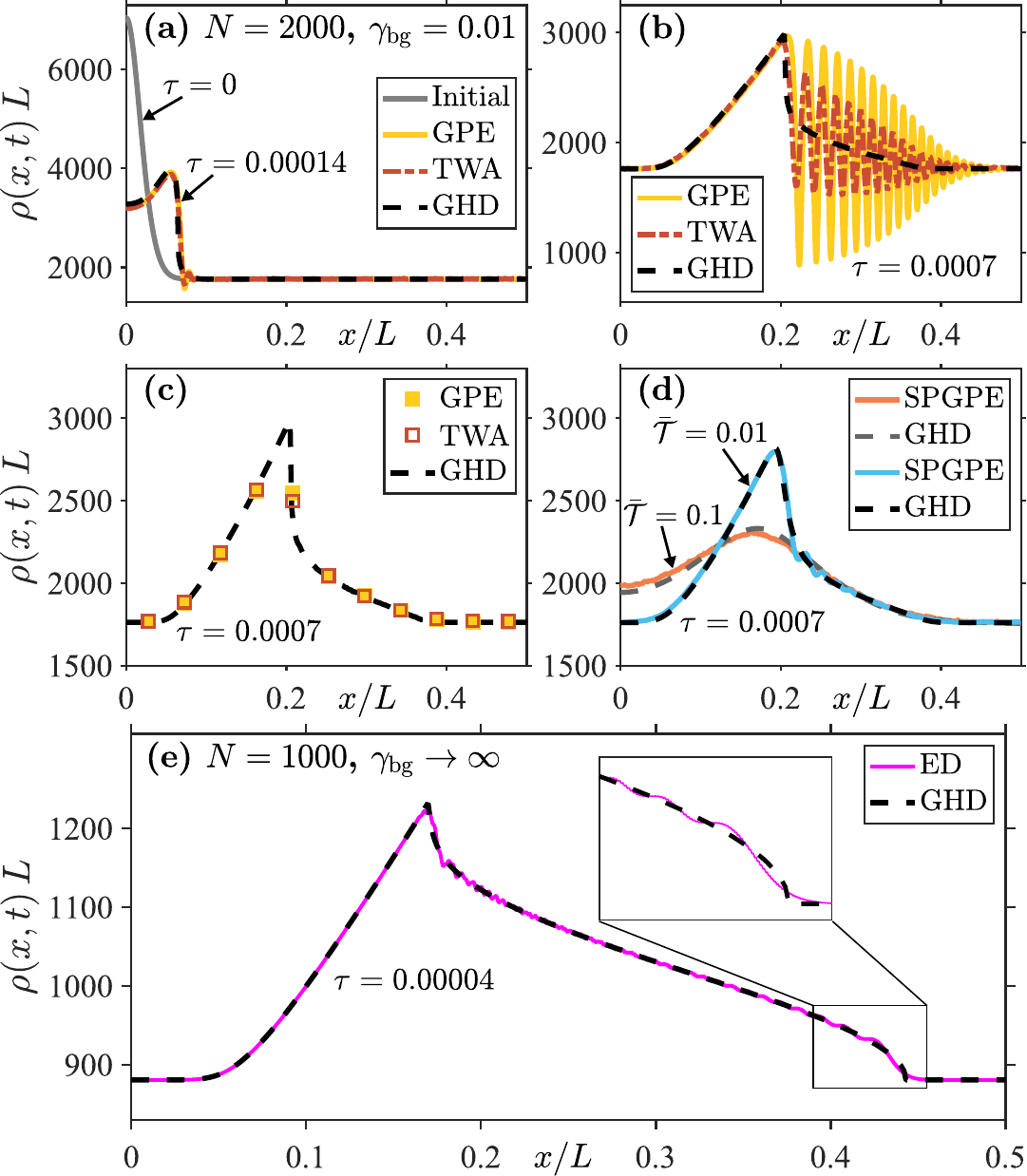}
    \caption{Dynamics of dispersive quantum shock waves forming in the evolution of an initial localized density bump in an otherwise uniform 1D Bose gas. Shown are snapshots of the evolving density profile at different dimensionless times $\tau=t\hbar/mL^2$, where $L$ is the length of the system, for different total atom numbers $N$ and dimensionless interaction strength $\gamma_{\mathrm{bg}}$ in the background. Due to the reflectional symmetry about the origin, we only show the densities for $x > 0$. See text for further discussion. Adapted from \cite{watson2022benchmarks}.
    }
    \label{fig:benchmark}
\end{figure}

Examples of simulations from Ref.~\cite{watson2022benchmarks} of dispersive quantum shock waves  initiated from dissolution of a localised density bump are shown in Figs.~\ref{fig:benchmark}\,(a)--(d) for a weakly interacting regime of the 1D Bose gas, $\gamma_{\mathrm{bg}}\ll 1$, and in Fig.~\ref{fig:benchmark}\,(e) in the Tonks-Girardeau
 regime of infinitely strong interactions ($\gamma_{\mathrm{bg}}\to \infty$). Here, $\gamma_{\mathrm{bg}}=mg/\hbar^2\rho_{\mathrm{bg}}$ is the dimensionless interaction strength at background density $\rho_{\mathrm{bg}}$. In (a) and (b), the predictions of GHD are compared with the mean-field Gross-Pitaevskii equation (GPE), and the truncated Wigner approach (TWA) which takes into account the effect of quantum fluctuations on top of the mean-filed description. In (c) the prediction of GHD at time $\tau=0.0007$ is compared with coarse-grained averages of the GPE and TWA results and shows good agreement. Such coarse-graining that involves convolution averaging ~\cite{watson2022benchmarks}, mimics finite imaging resolution in ultracold atom experiments and may explain the success of GHD when compared to experiments, even though its predictions may depart from other theoretical approaches that are valid at short wavelengths.

 In Fig.~\ref{fig:benchmark}\,(d), the GHD results at $\tau=0.0007$ are compared with the predictions of stochastic projected Gross-Pitaevskii equation (SPGPE) for a finite-temperature initial system, for two different initial dimensionless temperatures $\overline{\mathcal{T}}$ (where $\overline{\mathcal{T}}\!=\!T/T_d$, with $T_d\!=\!\hbar^2\rho_{\mathrm{bg}}^2/2mk_B$ being the temperature of quantum degeneracy), again showing excellent agreement. The performance of GHD generally improves with temperature as short-wavelength interference fringes present in an equivalent GPE ($T=0$) simulation get washed out by thermal fluctuations so that the microscopic dynamics better conform the scope of the GHD. Finally, in Fig.~\ref{fig:benchmark}(e) we compare a snapshot of the density profile at $\tau=0.00004$ (evolved from the same initial density bump) in the Tonks-Girardeau regime calculated using GHD and exact diagonalization (ED) of an equivalent free-fermion problem to which a Tonks-Girardeau gas can be mapped. Apart from missing density oscillations on short (microscopic) length scales, GHD agrees very well with the ED result in this strongly interacting regime as well.

\begin{figure}[tbp]
    \centering
    \includegraphics[width=8.2cm]{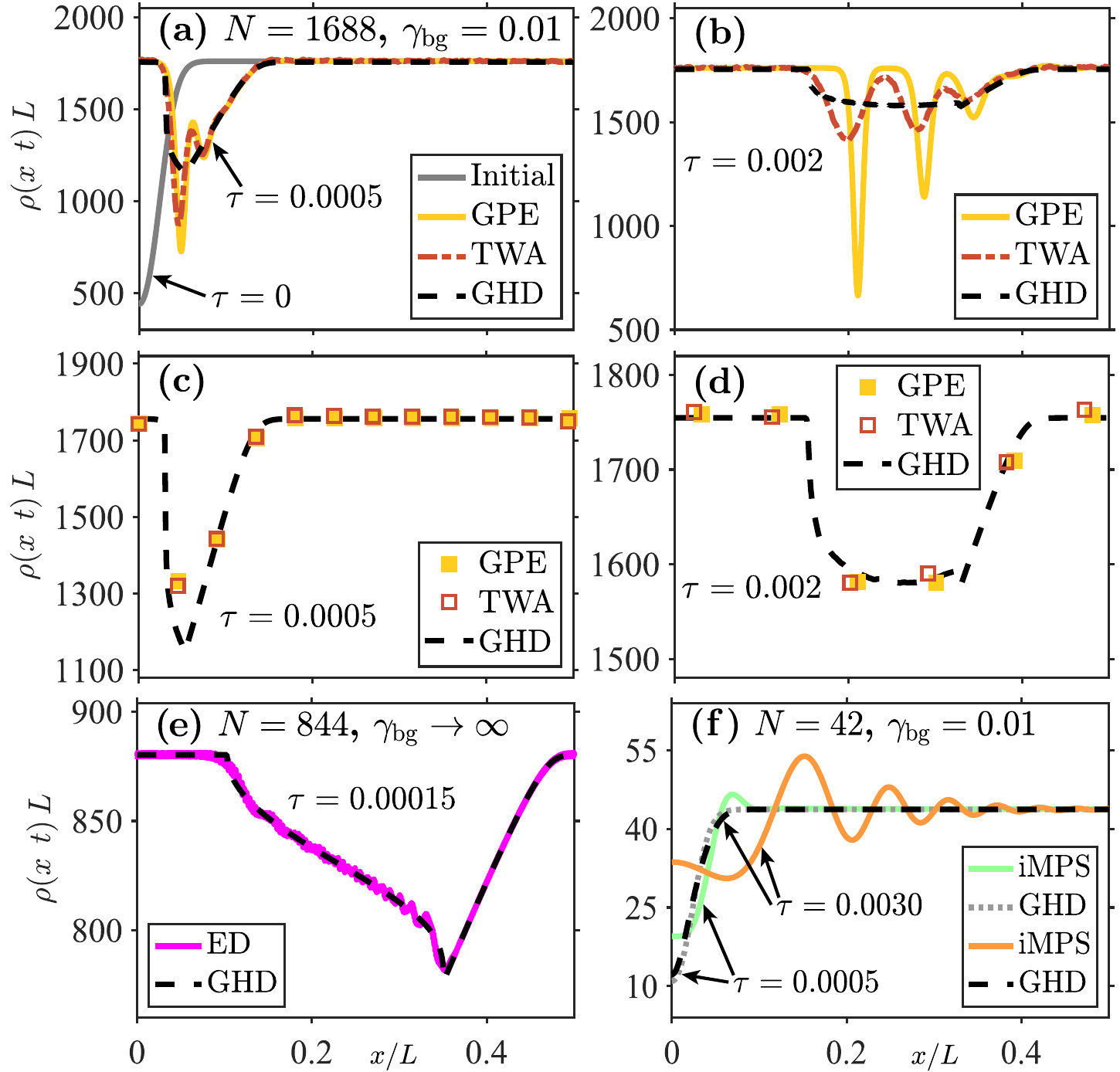}
    \caption{Same as in Fig.~\ref{fig:benchmark}, except for the evolution of an initial density dip.
    Panel (a) shows the initial density profile ($\tau=0$), and a snapshot of the time-evolved profiled at an early time ($\tau=0.0005$), evaluated using GPE, TWA, and GHD. Panel (b) is a snapshot of the density profile at a later time ($\tau=0.002$). A train of three gray solitons can be clearly identified in the mean-field GPE results; their visibility diminishes once quantum fluctuations are taken into account through the TWA approach. While GHD fails to capture individual soliton profiles or short-wavelength structures, it agrees well with the coarse-grained averages of the GPE and TWA results, shown in panels (c) and (d)  at $\tau=0.0005$ and $\tau=0.002$, respectively. Panel (e) is a comparison of GHD and ED results in the Tonks-Girardeau regime, showing excellent agreement. Finally, panel (f) shows a comparison of GHD with exact MPS calculations for $N=42$ in the nearly ideal Bose gas regime, where LDA is not applicable, hence explaining the disagreement of GHD with MPS results. Adapted from \cite{watson2022benchmarks}.}
    \label{fig:benchmark-dip}
\end{figure}

For intermediate interactions strengths, Watson \textit{et al.} \cite{watson2022benchmarks} have also compared the GHD results for the evolution of an initial bump with the results of the numerically exact MPS method for $N=50$ particles at zero temperature. They found that the performance of GHD generally improves with stronger interactions as short-wavelength interference phenomena become increasingly suppressed due to the reduced phase coherence in the system at stronger interactions.

Watson \emph{et al.} have also shown that  GHD fails to capture interference phenomena at short-length scales, even in coarse-grained convolution averaging sense, for very low temperatures and very weak interaction strengths, where the local density approximation, intrinsic to any hydrodynamic approach, also fails.

Similar conclusions have been obtained from simulations of an initial density dip (see Fig.~\ref{fig:benchmark-dip}), known to shed a train of gray solitons in the weakly interacting regime in the mean-field GPE approximation. Here, the situation is similar to that of Figs.~\ref{fig:benchmark}\,(b) and (c): while GHD fails to capture individual solitons, whose characteristic width (on the order of the microscopic healing length) lies beyond the intended range of applicability of any hydrodynamic theory, it adequately captures the coarse-grained average density over the soliton train.

In the benchmarks of GHD in the quantum Newton’s cradle setup, performed for a double-well to single-well trap quench of a weakly interacting quasicondensate, Watson \textit{et al.} \cite{watson2022benchmarks} observed excellent agreement with the SPGPE results in both the transient dynamics and the final relaxed state, as well as in the overall characteristic thermalization timescale. However, in the comparison for the quantum Newton's cradle initiated by Bragg pulses, they observed disagreement with SPGPE in the characteristic thermalization timescale \cite{Thomas2021Thermalization}, with the GHD generally predicting slower thermalization. This discrepancy, however, can be attributed to qualitatively very different ways that the Bragg pulses are implemented in GHD and the SPGPE (see Ref.~\cite{watson2022benchmarks} for further details).

\section{Conclusion and Perspectives}

Generalised hydrodynamics has recently emerged as a broadly applicable hydrodynamic theory for modelling the quantum many-body dynamics of integrable and near-integrable systems on a large scale. Since its inception, GHD has been extended to account for various experimentally relevant effects such as diffusion, dimensional crossover in the 1D Bose gas, inhomogeneous interactions, and quantum effects. Experimentally, tests of GHD are still in their infancy. However, the first few experimental investigations show excellent agreement between the observed results and GHD predictions.

The unreasonable effectiveness of GHD at low particle numbers in \cite{Malvania2021GHD} warrants further investigation, as do tunnelling effects between 1D tubes. Tunnelling between tubes breaks integrability and is therefore expected to lead to complete thermalization over sufficiently large time scales \cite{Bayocboc2022Dynamics,Panfil2023Thermalization}. Future experimental tests of GHD would benefit by exploring more challenging scenarios that push the assumptions of GHD, such as by straying further from integrability via long-range dipolar interactions. Experimental verification of diffusive GHD and hence characteristic thermalization rates \cite{Bastianello2020Thermalisation,Thomas2021Thermalization,watson2022benchmarks} in different regimes of the 1D Bose gas would also be an important achievement.

In the context of verifying the required conditions for GHD, we also mention the recent work by Le \textit{et al.} \cite{le2022direct},
which probed the rapid onset of hydrodynamics -- referred to as hydrodynamization -- in an array of 1D Bose gases in a strongly interacting regime. Hydrodynamization precedes local prethermalization, and it was conjectured that GHD can be applied immediately after hydrodynamization, i.e., during local prethermalization, even though the local GGE is not yet established. If confirmed, this conjecture would imply a further relaxation of the conditions required for the applicability of GHD, which in turn would explain the success of GHD under a broader range of experimental conditions.

While we have reviewed GHD in the context of a repulsive 1D Bose gas, GHD is also applicable to the attractive 1D Bose gas \cite{Koch2021Adiabatic,Koch_2022} and to other integrable models, such as the Hubbard model \cite{DeNardis2017Hubbard,Nozawa2020_GHD_Hubbard,Nozawa2021GHD_Hubbard}, the XXZ chain \cite{ghydro2,Piroli2017XXZ,Bulchandani2021Superdiffusion}, and the Yang-Gaudin model, describing the spin--$1/2$ Fermi gas \cite{gaudin1967systeme, scopa2022generalized, PhysRevLett.19.1312}. (For a recent review on advances in the study of transport in spin chains, sparked by the theory of GHD, we direct the reader to Ref.~\cite{Bulchandani2021Superdiffusion}.) As such, it would be beneficial to see experimental verification of GHD in these and other condensed matter systems. Finally, it would be interesting to use GHD to model the non-equilibrium dynamics of more general integrable systems such as the multi-component Bose gas \cite{PhysRevLett.95.150402} and Bose-Fermi mixtures \cite{PhysRevA.99.013628}. 

%\begin{acknowledgments}
%K.\,V.\,K. acknowledges support by the Australian Research Council Discovery Project Grant DP190101515.
%\end{acknowledgments}

~

\noindent \textbf{Acknowledgements}\\
The authors acknowledge stimulating discussions with R. S. Watson and G. De Rosi.

~

\noindent \textbf{Authors' contributions}\\
M.\,L.\,K.\, wrote the first draft; K.\,V.\,K. expanded and completed the writing of the manuscript. All authors proofread and approved the final version of the manuscript. 

~

\noindent \textbf{Funding}\\
This work was supported by the Australian Research Council Discovery Project Grant DP190101515.

~

\noindent \textbf{Availability of data and materials}\\
All data and figures presented in this review article are reproduced from published papers that are duly cited in figure captions.

~

\noindent \textbf{Competing interests}\\
The authors declare that they have no competing interests.

~

%\bibliography{Bibliography}

%apsrev4-2.bst 2019-01-14 (MD) hand-edited version of apsrev4-1.bst
%Control: key (0)
%Control: author (8) initials jnrlst
%Control: editor formatted (1) identically to author
%Control: production of article title (0) allowed
%Control: page (0) single
%Control: year (1) truncated
%Control: production of eprint (0) enabled
%

\end{document}